\documentclass{aa}

\usepackage{psfig}

\newcommand\tn{TN~J1338$-$1942}

\newcommand\cf{{c.f.,~}}
\newcommand\eg{{e.g.,~}}
\newcommand\etal{{et al.~}}
\newcommand\ie{{i.e.,~}}
\newcommand\Lya{Ly$\alpha$}
\newcommand\CIV{\hbox{C~$\rm IV$}~$\lambda$~1549}
\newcommand\HeII{\hbox{He~$\rm II$}~$\lambda$~1640}
\newcommand\HI{\ion{H}{i}}
\newcommand\HII{\ion{H}{ii}}

\newcommand\aap{{A\&A}}
\newcommand\aasup{{A\&AS}}
\newcommand\aj{{AJ}}
\newcommand\apj{{ApJ}}
\newcommand\apjl{{ApJ}}

\newcommand\mnras{{MNRAS}}
\newcommand\nature{{Nature}}

\def\spose#1{\hbox to 0pt{#1\hss}}
\newcommand\simlt{\mathrel{\spose{\lower 3pt\hbox{$\mathchar"218$}}
     \raise 2.0pt\hbox{$\mathchar"13C$}}}
\newcommand\simgt{\mathrel{\spose{\lower 3pt\hbox{$\mathchar"218$}}
     \raise 2.0pt\hbox{$\mathchar"13E$}}}

\begin{document}

\thesaurus{11.01.2; 11.09.1; 12.03.3}

\title{VLT Spectroscopy of the z=4.11 Radio Galaxy TN~J1338$-$1942\thanks{Based on observations at the ESO VLT Antu telescope}}

\subtitle{}

\author{Carlos De Breuck \inst{1,2} \and Wil van Breugel \inst{2} \and Dante Minniti \inst{3,2}  \and George Miley \inst{1} \and Huub R{\"o}ttgering \inst{1} \and S. A. Stanford \inst{2} \and Chris Carilli \inst{4}}

\offprints{Carlos De Breuck}

\institute{Sterrewacht Leiden, Postbus 9513, 2300 RA Leiden, The
             Netherlands; debreuck,miley,rottgeri@strw.leidenuniv.nl
             \and Institute of Geophysics and Planetary Physics, Lawrence Livermore National Laboratory, L-413, Livermore, CA 94550, U.S.A.; wil,adam@igpp.llnl.gov \and P. Universidad Catolica, Avda. Vicuna Mackenna 4860,
             Casilla 104, Santiago 22, Chile; dante@astro.puc.cl \and National Radio Astronomy Observatory, Socorro, NM 87801, USA; ccarilli@nrao.edu
             }

\date{Received 1999; accepted 1999}

\maketitle

\begin{abstract}

We present optical, infrared and radio data of the $z = 4.11$ radio
galaxy \tn\, including an intermediate resolution spectrum obtained with FORS1
on the VLT Antu telescope. \tn\ was the first $z > 4$ radio galaxy to
be discovered in the southern hemisphere and is one of the most
luminous \Lya\ objects in its class.  The \Lya\ and rest--frame optical
emission appear co--spatial with the brightest radio hotspot of this very
asymmetric radio source, suggesting extremely strong interaction with dense
ambient clouds.

The \Lya\ is spatially extended by $\sim$ 4\arcsec (30~kpc), has an enormous
rest--frame equivalent width, $W_{\lambda}^{rest}=210 \pm 50$~\AA, and has a
spectral profile that is very asymmetric with a deficit towards the
blue. We interpret this blue-ward asymmetry as being due to absorption
of the \Lya\ photons by cold gas in a turbulent halo surrounding the
radio galaxy and show that the required neutral hydrogen column density
must be in the range $3.5 - 13 \times 10^{19}$ cm$^{-2}$. The
two-dimensional spectrum indicates that the extent of the absorbing gas
is comparable (or even larger) than the 4\arcsec (30~kpc) \Lya\ emitting
region.

The VLT observations are sufficiently sensitive to detect the
continuum flux both blue-ward and red-ward of the \Lya\ emission,
allowing us to measure the \Lya\ forest continuum break (\Lya\
'discontinuity', $D_A$) and the Lyman limit. We measure a $D_A=0.37
\pm 0.1$, which is $\sim 0.2$ lower than the values found for quasars
at this redshift. We interpret this difference as possibly due to a
bias towards large $D_A$ introduced in high--redshift quasar samples
that are selected on the basis of specific optical colors.  If such a
bias would exist in optically selected quasars, -- and even in samples
of Lyman break galaxies --, then the space density of both classes of
object will be underestimated.  Furthermore, the average \HI\ column
density along cosmological lines of sight as determined using quasar
absorption lines would be overestimated. Because of their radio-based
selection, we argue that $z>4$ radio galaxies are excellent objects
for investigating $D_A$ statistics.


\end{abstract}


\section{Introduction}

Within standard Cold Dark Matter scenarios the formation of galaxies is
a hierarchical and biased process. Large galaxies are thought to be assembled
through the merging of smaller systems, and the most massive objects
will form in over--dense regions, which will eventually evolve into the
clusters of galaxies (\cite{kau99}). It is therefore important to find
and study the progenitors of the most massive galaxies at the highest
possible redshifts.

Radio sources are convenient beacons for pinpointing massive elliptical
galaxies, at least up to redshifts $z\sim 1$ (\cite{lil84};
\cite{best98a}). The near--infrared `Hubble' $K-z$ relation for such
galaxies appears to hold up to $z= 5.2$, despite large K--correction
effects and morphological changes (Lilly and Longair 1984; van Breugel
\etal 1998, 1999).  \nocite{lil84,wvb98,wvb99a} This suggests that
radio sources may be used to find massive galaxies and their likely
progenitors out to very high redshift.

While optical, `color--dropout' techniques have been successfully used
to find large numbers of 'normal' young galaxies (without dominant AGN)
at redshifts surpassing those of quasars and radio
galaxies(\cite{wey98}), the radio and near--infrared selection technique
has the additional advantage that it is unbiased with respect to the
amount of dust extinction. High redshift radio galaxies (HzRGs) are
therefore also important laboratories for studying the large amounts of
dust (\cite{dun94}; \cite{ivi98}) and molecular gas (\cite{pap99}),
which are observed to accompany the formation of the first forming
massive galaxies.

Using newly available, large radio surveys we have begun a systematic
search for $z>4$  HzRGs to be followed by more detailed studies of
selected objects.  In this Letter, we present deep intermediate
resolution VLT/FORS1 spectroscopy of \tn\ which, at $z = 4.11$, was
the first $z > 4$ radio galaxy discovered in the southern hemisphere
(\cite{deb99a}), and is one of the brightest and most luminous \Lya\
objects of its class.

In \S 2, we describe the discovery and previous observations of
\tn. In \S 3 we describe our VLT observations, and in \S 4 we
discuss some of the implications of our results.  Throughout this
paper we will assume $H_0 = 65$~km~s$^{-1}$Mpc$^{-1}$, $q_0$=0.15,
and $\Lambda=0$. At $z=4.11$, this implies a linear size scale of
7.5 ~kpc/arcsec.


\section{Source selection and previous observations}

The method we are using to find distant radio galaxies is based on the
empirical correlation between redshift and observed spectral index in
samples of low-frequency selected radio sources (\eg \cite{car99}).
Selecting radio sources with ultra steep spectra (USS) dramatically
increases the probability of pinpointing high-z radio galaxies, as
compared to observing radio galaxies with more common radio spectra.
This method, which can to a large extent be explained as a K-correction
induced by a curvature of the radio spectra, has been shown to be
extremely efficient (\eg \cite{cha90}; \cite{wvb99a}).

We constructed such a USS sample ($\alpha^{\rm 1.4 GHz}_{\rm 365
MHz} < -1.30 ;S_\nu \propto \nu^\alpha$; \cite{deb99b}),
consisting of 669 objects, using several radio catalogs which, in
the southern hemisphere, include the Texas 365~MHz catalog
(\cite{dou96}) and the NVSS 1.4~GHz catalog (\cite{con98}).

As part of our search--program we observed \tn\ ($\alpha^{\rm 1.4
GHz}_{\rm 365 MHz} = -1.31\pm0.07$) with the ESO 3.6m telescope in 1997
March and April (\cite{deb99a}). The radio source was first identified
by taking a 10 minute $R-$band image.  Followup spectroscopy then showed
the radio galaxy to be at a redshift of $z=4.13 \pm 0.02$, based on a
strong detection of \Lya, and weak confirming \CIV\ and \HeII.  At this
redshift its derived rest--frame low frequency (178~MHz) radio
luminosity is comparable to that of the most luminous 3CR sources.

\begin{figure}

\psfig{file=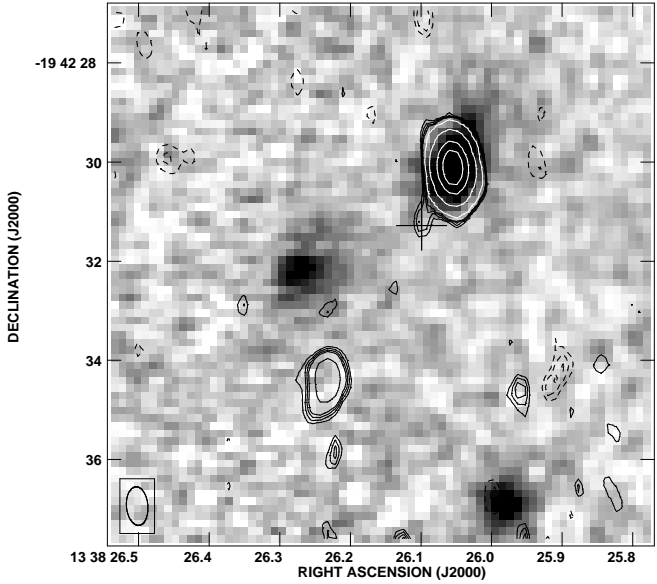,width=8.8cm,angle=-90}

\caption[]{4.85~GHz VLA radio contours overlaid on a Keck $K-$band
image. The cross indicates the position of the likely radio core at
8.5~GHz, which appears offset from the galaxy by $1\farcs4$ ($\sim 4
\sigma$) along the radio axis. Contour levels are $-$0.23, $-$0.17,
$-$0.12, 0.12, 0.15, 0.17, 0.20, 0.35, 1.45, 5.8, and 29 mJy/beam}

\label{kradio}

\end{figure}

More detailed radio information was obtained with the VLA at 4.71~GHz
and 8.46~GHz on 1998 March 24, as part of a survey to measure  rotation
measures in HzRGs (\cite{pen99}). We detect two radio components
($S_{4.7 GHz}^{NW} = 21.9$ mJy; $S_{4.7 GHz}^{SE} = 1.1$ mJy) separated
by 5\farcs5 in the field of the radio galaxy (Fig.  \ref{kradio}).  The
bright NW component has a very faint radio companion ($S_{4.7 GHz}^{C}
= 0.3$ mJy) at 1\farcs4 to the SE.  Our present observations show that
all components have very steep radio spectra with $\alpha^{\rm
8.5~GHz}_{\rm 4.7~GHz}(NW) \sim  -1.6$, $\alpha^{\rm 8.5~GHz}_{\rm
4.7~GHz}(SE) \sim  -1.8$, and $\alpha^{\rm 8.5~GHz}_{\rm 4.7~GHz}(C)
\sim  -1.0$. The proximity and alignment of such rare USS components
strongly suggests that they are related and part of one source. While
further observations over a wider frequency range would be useful to
confirm this, for now we conclude that \tn\ is a very asymmetric radio
source, and identify component C at $\alpha_{2000} = 13^h38^m26\fs10$
and $\delta_{2000}=-19\degr 42\arcmin 31\farcs1$ with the radio core.
Such asymmetric radio sources are not uncommon (\eg \cite{mcc91}), and
are usually thought to be due to strong interaction of one of its radio
lobes with very dense gas or a neighboring galaxy (see for example
\cite{fei99}).

We also obtained a $K-$band image with the Near Infrared Camera
(NIRC; \cite{mat94}) at the Keck I telescope on UT 1998 April 18.
The integration time was 64 minutes in photometric conditions with
0\farcs5 seeing. Observing procedures, calibration and data
reduction techniques were similar to those described in van
Breugel et al. (1998). Using a circular aperture of 3\arcsec,
encompassing the entire object, we measure $K=19.4 \pm 0.2$ (we do
not expect a significant contribution from emission lines at the
redshift of the galaxy). In a 64~kpc metric aperture, the
magnitude is $K_{64}=19.2 \pm 0.3$, which puts \tn\ at the bright
end, but within the scatter, of the $K-z$ relationship
(\cite{wvb98}).

We determined the astrometric positions in our $5\arcmin \times
5\arcmin$ $R-$band image using the USNO PMM catalog (\cite{mon98}). We
next used the positions of nine stars on the R-band image in
common with the Keck $K-$band to solve the astrometry on the $1\arcmin
\times 1\arcmin$ $K-$band image.  The error in the {\it relative}
near--IR/radio astrometry is dominated by the absolute uncertainty of
the optical reference frame, which is $\sim$0\farcs4 (90\% confidence
limit; \cite{deu99}).  
In figure \ref{kradio}, we show the overlay of the radio and $K-$band
({\it rest-frame} $B-$band) images. The NW hotspot coincides within
0\farcs035 of the peak of the $K-$band emission, while some faint
diffuse extensions can be seen towards the radio core and beyond the
lobe. The positional difference between the peak of the $K-$band
emission and the radio core is 1\farcs4 ($\sim 4\sigma$), which suggests
that the AGN and peaks of the $K-$band and \Lya\ emission may not be
co--centered.



\section{VLT observations}

Because of the importance of \tn\ as a southern laboratory for
studying HzRGs, we obtained a spectrum of this object with high
signal--to--noise and intermediate spectral resolution with FORS1 on
the ANTU unit of the VLT on UT 1999 April 20. The purpose of these
observations was to study the \Lya\ emission and UV--continuum in
detail.

The radio galaxy was detected in the acquisition images ($t_{int}=
2\times 60$~s; $I=23.0 \pm 0.5$ in a 2\arcsec\ aperture). We used the
600R grism with a 1\farcs3 wide slit, resulting in a spectral resolution
of 5.5~\AA\ (FWHM). The slit was centered on the peak of the $K-$band
emission at a position angle of 210\degr\ North through East.  To
minimize the effects of fringing in the red part of the CCD, we split
the observation into two 1400~s exposures, while offsetting the object
by 10\arcsec\ along the slit between the individual exposures. The
seeing during the \tn\ observations was $\sim$0\farcs7 and conditions
were photometric.

Data reduction followed the standard procedures using the NOAO IRAF
package.  We extracted the one-dimensional spectrum using a 4\arcsec\
wide aperture, chosen to include all of the \Lya\ emission. For the
initial wavelength calibration, we used exposures of a HeArNe lamp.  We
then adjusted the final zero point of the wavelength scale using telluric
emission lines. The flux calibration was based on observations of the
spectrophotometric standard star LTT2415, and is believed to be accurate
to $\sim 15\%$. We corrected the spectrum for foreground Galactic
extinction using a reddening of $E_{B-V}=0.096$ determined from the dust
maps of \cite{schl98}.



\begin{figure}

\psfig{file=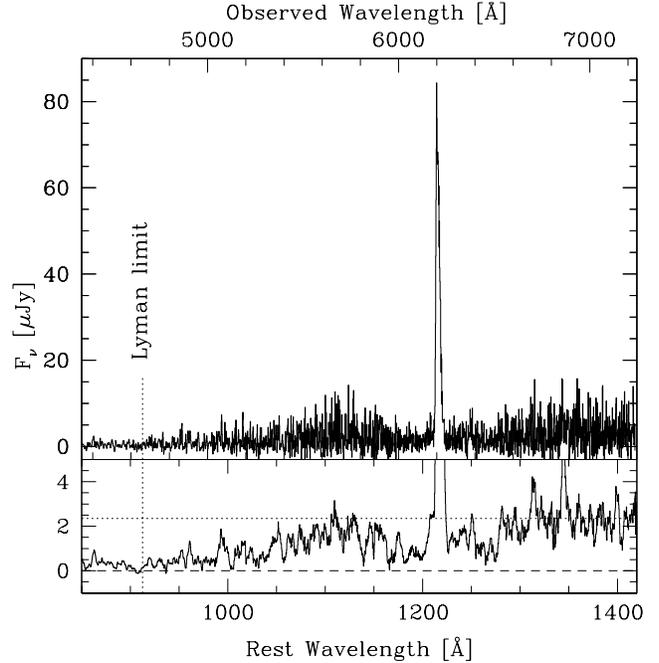,width=8.8cm}

\caption{VLT spectrum of \tn . The lower panel has been boxcar
smoothed by a factor of 15 to better show the shape of the \Lya\
forest and the Lyman limit. The horizontal dotted line is the
extrapolation of the continuum at 1300\AA\ $< \lambda_{rest} <
1400$\AA, and the vertical dotted line indicates the position of the
$\lambda_{rest}=912$\AA\ Lyman limit.}

\label{tn1338spec}

\end{figure}



\begin{figure}

\psfig{file=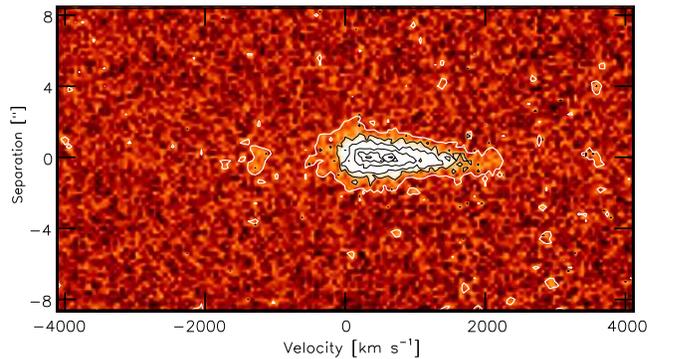,width=8.8cm}

\caption{Two dimensional FORS1 spectrum of the \Lya\ region. Note the
strong, 1400 km~s$^{-1}$ wide depression in the blue
half. \label{2DLya}}

\end{figure}

In figure \ref{tn1338spec} we show the observed one dimensional
spectrum and in figure \ref{2DLya} the region of the two-dimensional
spectrum surrounding the \Lya\ emission line. Most notable is the large  
asymmetry in the profile, consistent with a very wide ($\sim
1400$~km~s$^{-1}$) blue-ward depression. Following previous
detection of Ly $\alpha$ absorption systems in HzRGs (\cite{rot95}; 
\cite{oji97}; \cite{dey99}) we shall interpret the blue-ward asymmetry
in the \Lya\ profile of \tn\ as being due to foreground absorption by
neutral hydrogen.

The rest--frame equivalent width of \Lya\ in \tn\,
$W_{\lambda}^{rest}=210 \pm 50$~\AA, is twice as high as in the
well--studied radio galaxy 4C~41.17 ($z=3.80$; \cite{dey97}). The
large \Lya\ luminosity ($L_{\rm Ly\alpha} \sim 4\times 10^{44}$ erg
s$^{-1}$ after correction for absorption) makes \tn\ the most luminous
\Lya\ emitting radio galaxy known.

Following \cite{spi95}, we measure the continuum discontinuity
across the \Lya\ line, defined as [$\langle F_{\nu}(1250 -
1350{\rm \AA}) / F_{\nu}(1100 - 1200{\rm \AA}) \rangle$] = $1.56 \pm 0.24$.
Similarly, for the Lyman limit at $\lambda_{rest}$ = 912\AA, we
find [$\langle F_{\nu}(940 - 1000{\rm \AA}) / F_{\nu}(850 - 910{\rm \AA})
\rangle$] = $2.2 \pm 0.5$, though this value is uncertain because
the flux calibration at the edge of the spectrum is poorly
determined.

The presence of these continuum discontinuities further confirm our
measured redshift. However, the redshift of the system is difficult to
determine accurately because our VLT spectrum does not cover \CIV\ or
\HeII.  Furthermore, since the \Lya\ emission is heavily absorbed, it
is likely that the redshift of the peak of the \Lya\ emission (at $6206
\pm 4 {\rm \AA}$, $z=4.105\pm 0.005$) does not exactly 
coincide with the redshift of the galaxy. We shall assume $z = 4.11$.

\section{Discussion}

\tn\ shares several properties in common with other HzRGs but some of
its characteristics deserve special comment. Here
we shall briefly discuss these.



\begin{figure}
\psfig{file=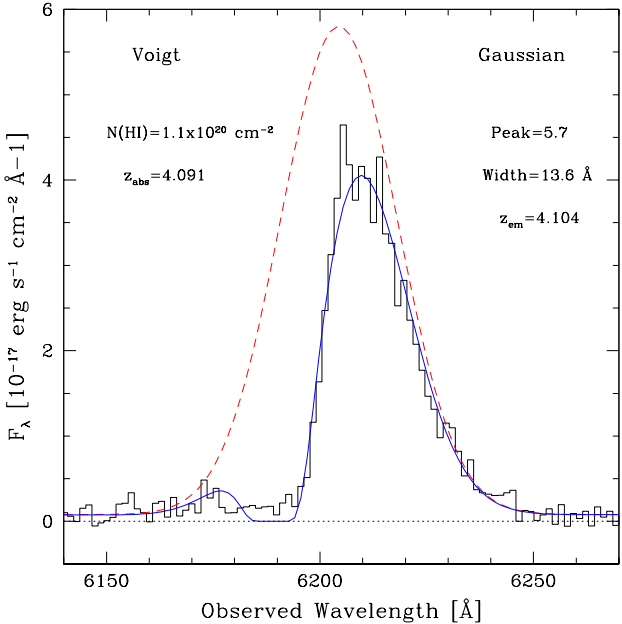,width=8.8cm}
\caption{Part of the spectrum around the \Lya\ line. The solid line is
the model consisting of a Gaussian emission profile (dashed line) and
a Voigt absorption profile with the indicated parameters.}
\label{lyamodel}

\end{figure}

\subsection{\Lya\ emission}

Assuming photoionization, case B recombination, and a temperature of
$T=10^4$~K we use the observed \Lya\ emission to derive a total mass
($M(\HII)$) of the \HII\ gas (\eg \cite{mcc90}) using
$M(\HII)=10^9(f_{-5}L_{44}V_{70})^{1/2}$~M$_{\odot}$. Here $f_{-5}$ is
the filling factor in units of 10$^{-5}$, $L_{44}$ is the \Lya\
luminosity in units of $10^{44}$~ergs~s$^{-1}$, and $V_{70}$ is the
total volume in units of $10^{70}$~cm$^3$.  Assuming a filling factor of
10$^{-5}$ (\cite{mcc90}), and a cubical volume with a side of 15~kpc, we
find $M(\HII) \approx 2.5\times 10^8$~M$_{\odot}$.  This value is on the
high side, but well within the range that has been found for HzRGs (\eg
\cite{oji97})).

Previous authors have shown that gas clouds of such mass can cause radio
jets to bend and decollimate (\eg \cite{wvb85,lon86,bar88}). Likewise,
the extreme asymmetry in the \tn\ radio source could well be the result
of strong interaction between the radio--emitting plasma and the \Lya\
gas.

\subsection{\Lya\ absorption}

Our spectrum also shows evidence for deep blue-ward absorption of the
\Lya\ emission line.  We believe that this is probably due to resonant
scattering by cold \HI\ gas in a halo surrounding the radio galaxy, as
seen in many other HzRGs (\cf \cite{rot95}, \cite{oji97},
\cite{dey99}).  The spatial extent of the absorption edge as seen in
the 2-dimensional spectrum (Fig. 3) implies that the extent of the
absorbing gas is similar or even larger than the 4\arcsec (30~kpc)
\Lya\ emitting region.

To constrain the absorption parameters we constructed a simple model
that describes the \Lya\ profile with a Gaussian emission function and a
single Voigt absorption function. As a first step, we fitted the red
wing of the emission line with a Gaussian emission profile. Because the
absorption is very broad, and extends to the red side of the peak, the
parameters of this Gaussian emission profile are not well
constrained. We adopted the Gaussian that best fits the lower red wing
as well as the faint secondary peak, 1400 km s$^{-1}$ blue-wards from
the main peak. The second step consisted of adjusting the parameters of
the Voigt absorption profile to best match the sharp rise towards the
main peak. The resulting model (shown along with the parameters of both
components in figure \ref{lyamodel}) adequately matches the main
features in the profile.  We varied the parameters of both components,
and all acceptable models yield column densities in the range $3.5
\times 10^{19} - 1.3 \times 10^{20}$ cm$^{-2}$.

The main difference between our simple model and the observations is the
relatively flat, but non--zero flux at the bottom of the broad
depression. This flux is higher than the continuum surrounding the \Lya\
line, indicating some photons can go through (\ie a filling factor less
than unity) or around the absorbing cloud.  If the angular size of
absorber and emitter are similar, the size of the absorber is $R_{abs}$
$\sim$10~kpc. The total mass of neutral hydrogen then is $2 - 10 \times
10^7 M_{\odot}$, comparable to or somewhat less than the total mass of
\HII.

\subsection{Continuum}

Following \cite{dey97}, and assuming that the rest frame UV continuum
is due to young stars, one can estimate the star--formation rate (SFR)
in \tn\ from the observed rest--frame UV continuum near 1400~\AA. From
our spectrum we estimate that $F_{1400} \sim 2 \mu Jy$, resulting in a
UV luminosity $L_{1400~{\rm \AA}} \sim 1.3 \times 10^{42}$
erg~s$^{-1}$~\AA$^{-1}$ and implying a SFR between 90 $-$ 720
h$_{65}^{-2}$ M$_{\odot}$ yr$^{-1}$ in a $10 \times 30$~kpc$^2$
aperture. These values are similar to those found for 4C~41.17. In
this case detailed HST images, when compared with high resolution
radio maps, strongly suggested that this large SFR might have been
induced at least in part by powerful jets interacting with massive,
dense clouds (\cite{dey97}; \cite{wvb99b}; \cite{bic99}).  The
co--spatial \Lya\ emission--line and rest--frame optical continuum
with the brightest radio hotspot in \tn\ suggests that a similar
strong interaction might occur in this very asymmetric radio source.

The decrement of the continuum blue-wards of \Lya\ (Fig.
\ref{tn1338spec}) due to the intervening \HI\ absorption along the
cosmological line of sight is described by the ``flux deficit''
parameter $D_A=\langle
1-{{f_{\nu}(\lambda1050-1170)_{obs}}\over{f_{\nu}(\lambda1050-1170)_{pred}}}
\rangle$ (\cite{oke82}). For \tn\ we measure $D_A=0.37 \pm 0.1$,
comparable to the $D_A=0.45 \pm 0.1$ that \cite{spi95} found for the
$z=4.25$ radio galaxy 8C~1435+64 (uncorrected for Galactic
reddening). This is only the second time the $D_A$ parameter has been
measured in a radio galaxy.

The decrement described by $D_A$ is considered to be extrinsic to the
object toward which it has been measured, and should therefore give
similar values for different classes of objects at the same
redshift. Because they have bright continua, quasars have historically
been the most popular objects to measure $D_A$. For $z \sim 4.1$,
quasars have measured values of $D_A \sim 0.55$ (\eg Schneider,
Schmidt \& Gunn 1991, 1997 \nocite{schn91,schn97}). Similar
measurements for color selected Lyman break galaxies do not yet exist.

Other non-color selected objects, in addition to radio galaxies, which
do have reported $D_A$ measurements are serendiptiously discovered
galaxies ($z=5.34, D_A > 0.70$, \cite{dey98}) and narrow-band
\Lya-selected galaxies ($z=5.74, D_A=0.79$, \cite{hu99}). Because of
their larger redshifts these galaxy values can not directly be
compared with those of quasars ($z_{max}=5.0, D_A=0.75$,
\cite{son99}).  However, they seem to fall slightly ($\Delta D_A \sim
-0.1$) below the theoretical extrapolation of Madau (1995) at their
respective redshifts, which quasars do follow rather closely.  This is
also true for the two radio galaxies ($\Delta D_A \sim -0.2$) at their
redshifts.  Thus it appears that non-color selected galaxies, whether
radio selected or otherwise, have $D_A$ values which fall below those
of quasars.

Although, with only two measurements, the statistical significance of
the low radio galaxy $D_A$ values is marginal, the result is
suggestive. It is worthwhile contemplating the implications that would
follow if further observations of $z > 4$ radio galaxies and other
objects selected without an optical color bias confirmed this
trend. Given that optical color selection methods (often used to find
quasars, and Lyman break galaxies) favour objects with large $D_A$
values, it is perhaps not surprising that non-color selected $z>4$
objects might have lower values of $D_A$.  Consequently, quasars and
galaxies with low $D_A$ values might be missed in color--based
surveys.  This then could lead to an underestimate of their space
densities, and an overestimate of the average \HI\ columns density
through the universe.

Radio galaxies have an extra advantage over radio selected quasars (\eg
\cite{hoo98}), because they very rarely contain BAL systems (there is
only one such example, 6C~1908+722 at $z=3.537$; \cite{dey99}). Such BAL
systems are known to lead to relatively large values of $D_A$, indicating
that part of the absorption is not due to cosmological HI gas, but due
to absorption within the BAL system (\cite{oke82}).  A statistically
significant sample of $z>4$ radio galaxies would therefore determine the
true space density of intervening \HI\ absorbers.

\section{Conclusions}
Because of its enormous \Lya\ luminosity and strong continuum, its
highly asymmetric and broad \Lya\ profile, and its very asymmetric
radio/near--IR morphology \tn\ is a unique laboratory for studying the
nature of $z > 4$ HzRGs. It is particularly important to investigate
the statistical properties of similar objects by extending the work
begun here to a significant sample of $z > 4$ HzRGs. The VLT will be a
crucial facility in such a study.

\begin{acknowledgements}

We thank the referee, Hy Spinrad, for his comments, which have
improved the paper. We also thank Remco Slijkhuis for his help in
using the ESO archive, and M\~y H\`a Vuong for useful discussions. The
W. M. Keck observatory is a scientific partnership between the
University of California and the California Institute of Technology,
made possible by the generous gift of the W. M. Keck Foundation. The
National Radio Astronomy Observatory is operated by Associated
Universities Inc., under cooperative agreement with the National
Science Foundation. The work by C.D.B., W.v.B., D.M.\ and S.A.S.\ at
IGPP/LLNL was performed under the auspices of the US Department of
Energy under contract W-7405-ENG-48. DM is also supported by Fondecyt
grant No. 01990440 and DIPUC.

\end{acknowledgements}

\end{document}